\documentclass[12pt]{iopart}
\usepackage{iopams}  
\usepackage{graphicx} 
\usepackage{xcolor}
\usepackage{multirow}

\date{May 2024}

\begin{document}

\title[]{Quantum cryptography visualized: \\ assessing visual attention on multiple representations with eye tracking in an AR-enhanced quantum cryptography student experiment}

\author{David Dzsotjan$^1$, Atakan Coban$^1$, Christoph Hoyer$^1$, Stefan K\"{u}chemann$^1$, J\"{u}rgen Durst$^2$, Anna Donhauser$^1$, Jochen Kuhn$^1$}

\address{$^1$ Chair of Physics Education, Faculty of Physics, Ludwig-Maximilians-Universität M\"{u}nchen (LMU) \\
$^2$ Physics Laboratory Courses, Faculty of Physics, Ludwig-Maximilians-Universität M\"{u}nchen (LMU)
}
\ead{david.dzsotjans@lmu.de}
\vspace{10pt}
\begin{indented}
\item[]October 2024
\end{indented}

\begin{abstract}
With the advent and development of real-world quantum technology applications, a practically-focused quantum education including student quantum experiments are gaining increasing importance in physics curricula. In this paper, using the DeFT framework, we present an analysis of the representations in our AR-enhanced quantum cryptography student experiment, in order to assess their potential for promoting conceptual learning. We also discuss learner visual attention with respect to the provided multiple representations based on the eye gaze data we have obtained from a pilot study where N=38 students carried out the tasks in our AR-enhanced quantum cryptography student experiment.
    
\end{abstract}

\section{Introduction}

\subsection{Quantum physics lab courses in physics curricula: an emerging phenomenon}


Laboratory physics courses are an integral part of physics curricula in university education. They promote learning through hands-on experience, helping students to acquire the skills of preparing and operating an experimental setup, to collect and interpret data, as well as to present their findings \cite{Wilcox2017}.


In recent years quantum physics  has been gaining more and more prominence and promise in real-world applications outside of academia, and thus several quantum education programs have been created to equip students with an application-oriented knowledge in quantum physics \cite{Asfaw2022, perron2021, Meyer2022}. The majority of the work has so far been focused on developing lecture-based courses  \cite{Aiello2021, Kaur2022, Singh2008}. Including quantum physics lab courses would most likely have further benefits in having a hands-on experience in working with quantum technologies \cite{Borish2022} and, so far, there are a few works investigating best practices for designing quantum experiments that promote conceptual learning \cite{Borish2023, Pearson2010, Galvez2010, Lukishova2022}.  Single-photon experiments are especially suited to improve the conceptual understanding of students about quantum physics \cite{Bitzenbauer2022}, for example: violation of Bell's inequality \cite{Dehlinger2002, Carlson2006}, single-photon interference through a double slit \cite{Lukishova2022}, Hong-Ou-Mandel interference \cite{Carvioto-Lagos2012} or quantum key distribution \cite{Bista2021}.

\subsection{DeFT and learning with Multiple External Representations}

 However, "merely" providing an experiment for students to carry out will not necessarily promote conceptual learning. The type of lab course where a succession of steps is defined as tasks, is not particularly well-suited for this \cite{Wieman2015, Holmes2017, Holmes2018}. Instead, if an inquiry-based learning style is encouraged where the learners are required to reflect on their actions during the experiment, conceptual learning is much more likely \cite{Etkina2010, Kontro2018}. Of course, designing such a learning environment is also a more challenging process. 

Key elements of learning environments are the representations that provide information needed for understanding a concept and/or for executing the tasks at hand. Since learners often differ in ways of absorbing information and constructing mental models, having not one but multiple representations for the same concept  can be an effective way to promote conceptual learning \cite{Ainsworth2006, Rexigel2024} As explained in \cite{Ainsworth2006}, the DeFT (Design, Functions, Tasks) framework presented therein can be used to gain a clear understanding about how multiple external representations (MER) in a learning environment contribute to the learners' progress. 

Within this framework, the key functions of MER for promoting conceptual understanding are threefold. Firstly, they can have a \emph{complementary function} either by containing \emph{complementary information} about a given concept or by supporting \emph{complementary processes}. The latter of these can manifest as choosing the most convenient representation based on \emph{individual differences} in learners, on the suitability of the provided information to solve the current \emph{task} or on the \emph{strategy} prompted by the individual representations. Secondly, MER can have a \emph{constraining function}, the interpretation of one representation being constrained by another. Thus, proceeding from the more general to the more particular representation the learner can arrive at  more precise interpretation. Lastly, MER may have a \emph{constructing function} where, through the process of \emph{abstraction}, \emph{knowledge extension} or \emph{relational understanding} between representations, the learner may arrive at a deeper understanding of the concept than by using only a single representation.

Among the different fields of physics, quantum mechanics includes one of the most abstract and hard to understand concepts: it is often considered very mathematical \cite{Johansson2018, Johansson2018_2, Corsiglia2020} and not related to the real world \cite{Hoehn2017, Dreyfus2019}. Thus, when designing learning environments for quantum physics curricula in university education, the DeFT framework is an especially valuable tool for making sure that the MER of the given environment do promote conceptual learning.


\subsection{AR-enhanced experiments and conceptual learning}

Augmented reality (AR) is a technology that can be used for designing learning environments that promote conceptual learning through inquiry-based experimentation \cite{Vidak2022, Ibanez2018}. Using AR, one can add a virtual layer with helpful visualizations to an existing experimental setup, introducing single or even multiple external representations.



So far, however, the majority of AR-enhanced physics student experiments use this technology to simulate parts of the original experimental setup \cite{Ibanez2014, Teichrew2020, Akcayir2016}. This can be a useful approach for hybrid experiments where the simulated units would be too dangerous to use or very expensive to procure. In a few cases, however, the AR-enhancement is applied to a fully functional experimental setup, and introducing virtual elements that display measurement data \cite{Strzys2018, Altmeyer2020}, and also react to motoric interactions with the experimental setup in real time \cite{Schlummer2023}. In these cases, the task of the AR-enhancement is purely to promote conceptual learning, rather than to simulate parts of the experiment.

\subsection{An AR-enhanced quantum cryptography experiment with eye tracking}

In this paper we present our AR-enhanced quantum cryptography student experiment and use the DeFT framework to discuss the functions of the MER within the learning environment. Using the learners' eye gaze data, we also perform an inferential analysis to understand their attention distribution among the MER during the experiment.




The paper is structured as follows. Sec. ~\ref{sec:QKD experiment} gives an overview about the theory and the experimental setup. We give a brief description about quantum key distribution (QKD) and the underlying key quantum physical features, discuss the basic experimental setup, and discuss the visualizations we use in the AR-enhanced experiment. 

In Sec. ~\ref{sec:eyegaze_analysis} we present the tasks to be carried out in the student experiment we used in the pilot study, and take a look at the eye gaze data in terms of fixations and use it to quantify the learners' attention to the virtual and real objects in the learning environment. Finally, in Sec. ~\ref{sec:conclusion}, we summarize our findings.

\section{Theoretical background and MER in the experiment} \label{sec:QKD experiment}

    \subsection{Quantum key distribution and the underlying physics}

    The AR-enhanced experiment described in this paper aims for students to understand the principle and use of the BB84 quantum key distribution protocol in a hands-on, inquiry-based setting. 
    
    When sending an encrypted message between two parties, two things have vital importance. Firstly, both parties should possess the same key with which the message can be encrypted and decrypted. Secondly, the transmission must be secure, i.e., a potential eavesdropper should not be able to access the encryption key without being noticed. Quantum cryptography, through the exploitation of quantum physics phenomena, imbues classical cryptography methods with a dramatically increased security for distributing the encryption key between the sending and receiving party.

    \subsubsection{One-time pad method}

The one-time pad method, otherwise known as the  singe-use key method, is a classical encryption protocol. It is in principle perfectly secure, provided both the sending and receiving parties have the same random encryption key, and the key is newly generated every time a new message is sent. 

Suppose Alice wants to send a message to Bob where the message is a string of characters in ASCII binary encoding. Also, let both parties have the same, randomly generated string of classical bits, i.e., 1-s and 0-s, of the same length as the message. The encrypted message is generated by binary addition of the message and the encryption key. Subsequently, Bob receives this encrypted message and decodes it by adding the encryption key to it once again, thus recovering the original message.

A crucial point for the security of the transmission is the distribution of the encryption key between the parties. One has to make sure that this key is not intercepted or, if there is an eavesdropper, their presence be reliably detected. As we will see in the following, the use of quantum bits, or qubits, will be instrumental in ensuring a secure encryption key distribution.

\subsubsection{Qubits and essential quantum features}

Let us regard two orthogonal states of a quantum system which we will call $|0\rangle$ and $|1\rangle$. If, by some suitable interaction, we can flip the state of the system between these two states, we can call it a qubit, or quantum bit, and its properties will include those of a classical bit. However, because of its vastly different physical realization, a qubit can do much more than a classical bit. Here, we will discuss its features relevant to quantum cryptography: coherent superposition, measurement, and the no-cloning theorem.

Let us choose $|0\rangle$ and $|1\rangle$ as an orthonormal basis so that they span a 2D Hilbert space:
\begin{eqnarray}
        \langle i|j\rangle = \delta_{i, j}, \quad i, j \in \{0, 1\},\\
\end{eqnarray}
where  $\delta_{i,j}$ is the Kronecker delta.

\begin{itemize}

\item Coherent superposition of states

Possible states of the qubit comprise $|0\rangle$, $|1\rangle$ and also their coherent superposition:
\begin{eqnarray}
    |\psi\rangle = a_0 |0\rangle + a_1 |1\rangle, \\
    |a_0|^2 + |a_1|^2 = 1,
\end{eqnarray}
where $a_0$ and $a_1$ are the linear combination coefficients or probability amplitudes. This is a state that cannot be found in the realm of classical physics. Its coherent nature refers to the fact that  $a_{0,1} \in \mathbb{C}$, with well-defined phases. 

In Quantum Key Distribution (QKD) this allows for working with the states of one basis and be able to express them as the coherent superpositions of states of another basis. This is essential in QKD because the use of 2 bases ensures that not only can the encryption key be shared but also a potential eavesdropper can be intercepted.

\item Qubit measurement

Another feature of a quantum system is its behavior when we perform a measurement on it. Let us suppose that we have a measuring device that measures whether the qubit is in state $|0\rangle$ or state $|1\rangle$ while the initial state of the qubit is a coherent superposition. Upon performing the measurement, the device will stochastically show either one or the other state as a result. Furthermore, the qubit's quantum state will simultaneously collapse into this measured state, so that subsequent repeated measurements on this same qubit will always result in the same state. 

We have stated that the outcome of the measurement is stochastic. If we have a detector in our measuring device that clicks if the incoming quantum state is $|i\rangle$, then, for a measurement of $|\psi\rangle$, the probability that this detector clicks is 
\begin{equation} \label{eq:measurement}
    \langle \psi| \left( |i\rangle\langle i| \right) |\psi \rangle = |a_i|^2, \quad i \in \{0, 1\},
\end{equation}
where $|i\rangle\langle i|$ is the projection operator representing the detection of state $|i\rangle$.

The stochastic nature introduced by the superposition state's projection onto one or the other measured basis state makes it possible to notice the presence of a potential eavesdropper by examining the statistical trends in the qubits transmitted between Alice and Bob.

\item No-cloning theorem

 In case of classical bits, we can measure the given bit and make a perfect copy of it, obtaining two identical bits. For a qubit, however, according to the no-cloning theorem, one cannot create an qubit identical to the original without destroying it. 

The no-cloning theorem prevents an eavesdropper to mask their presence by keeping the intercepted qubit for analysis and at the same time sending a perfect copy to Bob.

\end{itemize}

These three essential features underpin the quantum key distribution protocol, allowing for its high security. In our AR-enhanced experiment an important aim was to highlight them through the visualizations to prompt students to understand the underlying physical processes.

\subsubsection{BB84 protocol and quantum key distribution} \label{sec: QKD}

The BB84 quantum key distribution protocol is basically the one-time pad method with qubits and a twist: instead of only using bit values, we introduce the concept of bases. Let us choose two orthonormal bases that span the same Hilbert space. We will call them bases $+$ and $X$ and denote their respective states as $\{|0\rangle_k, |1\rangle_k\}$ where $k \in \{+, X\}$. Basis $X$ is rotated by 45 degrees compared to $+$, i.e.,
\begin{eqnarray}
        |0\rangle_X &=& \frac{1}{\sqrt{2}} \left(|0\rangle_+ - |1\rangle_+\right)\\
        |1\rangle_X &=& \frac{1}{\sqrt{2}} \left(|0\rangle_+ + |1\rangle_+\right).
\end{eqnarray}
We suppose that Alice on her end can generate all four quantum states at will and transmit them to Bob. He, in turn, can switch his measuring device so that it either measures qubits in the $+$ or in the $X$ basis.

All these provided, the quantum key distribution protocol goes as follows. Alice sends a string of qubits, randomly adjusting the basis ($+$ and $X$) and the bit value ($0$ and $1$). Bob, also randomly switching his detectors between $+$ and $X$ mode, measures them. Table ~\ref{tab:AB_transmission} lists the cases of basis choices by Alice and Bob and the likelihood that Bob measures the same bit value that Alice has sent. If the bases do not match then, in accordance with Eq. ~\ref{eq:measurement}, the qubit sent by Alice is randomly projected on one of the basis states at Bob.

\begin{table}[h!]
\centering
\begin{tabular}{|l|l|l|}
\hline
Basis by Alice & Basis by Bob & Probability of bit value matching \\ \hline
+              & +            & 100\%                            \\ \hline
+              & X            & 50\%                             \\ \hline
X              & +            & 50\%                             \\ \hline
X              & X            & 100\%                            \\ \hline
\end{tabular}
\caption{Probability that Bob measures the same bit value as what has been sent by Alice. If both bases are the same, then it is guaranteed that the measured bit value is correct, otherwise on average only half the instances are measured correctly.}
\label{tab:AB_transmission}
\end{table}
After Alice has finished sending qubits for generating the quantum key, both Alice and Bob will have a list of bases and related qubit values. Through a classical communication channel they subsequently compare their basis choices but \emph{not} the bit values. The instances where the bases did not match are discarded. They can thus be sure that for remaining instances the bit values for Alice and Bob are the same. These remaining bit values will then constitute the encryption key.

\subsubsection{QKD security: detecting an eavesdropper} \label{sec:detect_Eve}

The real advantage of quantum key distribution (QKD) is that it is principally impossible for an eavesdropper to intercept the shared key unnoticed. 

Let us suppose that a third person, Eve, is trying to tap the communication. Ideally, what she would want to do is make a perfect copy of the qubit that Alice has sent, measure the original qubit and pass the copy on to Bob in order to hide her presence. 

However, due to the no-cloning theorem, it would be impossible for her to make this perfect copy without destroying the original qubit. The next best thing she can do is to randomly adjust the basis in which she measures the incoming qubit, then generate a qubit in the state she has measured and send it further to Bob. 

As per the BB84 protocol, having sent the qubits from which the key is later constructed, Alice and Bob compare their basis choices and discard the transmission instances where their bases were different. In the absence of an eavesdropper they can be sure that the qubit values for the remaining instances are the same. However, if someone is tapping their communication, the situation is quite different, as Table  \ref{tab:AEB_transmission} shows.
\begin{table}[h!]
\centering
\begin{tabular}{|l|l|l|}
\hline
Bases by Alice and Bob & Basis by Eve & Probability of bit value matching \\ \hline
++              & +            & 100\%                            \\ \hline
++              & X            & 50\%                             \\ \hline
XX              & +            & 50\%                             \\ \hline
XX              & X            & 100\%                            \\ \hline
\end{tabular}
\caption{Probabilities that Bob measures the same bit value as what Alice has sent, in the presence of an eavesdropper (Eve), provided the bases of Alice and Bob are the same. If Eve's basis is different from those of Alice and Bob, there is a 50\% chance that Bob measures a different bit value than what Alice has sent.}
\label{tab:AEB_transmission}
\end{table}
For instances where the bases of Alice and Bob matched but the basis of Eve was different, there is only a $0.5$ probability that Bob measured the same bit value that Alice had sent. The total probability of bit value mismatch between Alice and Bob is
\begin{equation}
    \mathrm{P}_{\mathrm{mismatch}} = 0.5 \mathrm{P}_{++X} + 0.5\mathrm{P}_{XX+} ,
\end{equation}
where $\mathrm{P}_{++X}$ and $\mathrm{P}_{XX+}$ are probabilities for a basis choice of ++X and XX+, respectively, the subscript order referring to the bases of Alice, Bob and Eve. Supposing all basis choices were random, each basis combination shown in Table ~\ref{tab:AEB_transmission} has a probability of 0.25, provided we only take the instances where the bases of Alice and Bob were the same. Thus, $\mathrm{P}_{++X} = \mathrm{P}_{XX+} = 0.25$ which, in turn, results in $\mathrm{P}_{\mathrm{mismatch}} = 0.25$.

Thus, as the final step, Alice and Bob take the transmission instances where their bases coincided and publicly compare the bit values of a part of these measurements. If there is a mismatch of about 25\% then they can be certain that the quantum key transmission has been intercepted.

\subsection{Visualizations in the AR-enhanced experiment and their key functions in the DeFT framework}


The AR-enhancement in the experiment provides MER for the state of the $\lambda/2$ plates and for the qubit state at different points along the optical path. The goal behind these virtual objects is to promote engagement and immersion, as well as to encourage students to experiment with different $\lambda/2$ plate angles and to observe how the qubit states along the path of transmission change, thereby promoting a deeper understanding of the underlying physics. 

\subsubsection{$\lambda/2$ plates}

There are two kinds of $\lambda/2$ plates in the experiment. The first type has 4 different angle settings to create qubit states of bases + and X, respectively. The second has only 2 angle settings, allowing for qubit measurement in the + and X basis, respectively. 

\begin{figure}[h!]
    \centering
    \includegraphics[width=0.42\textwidth]{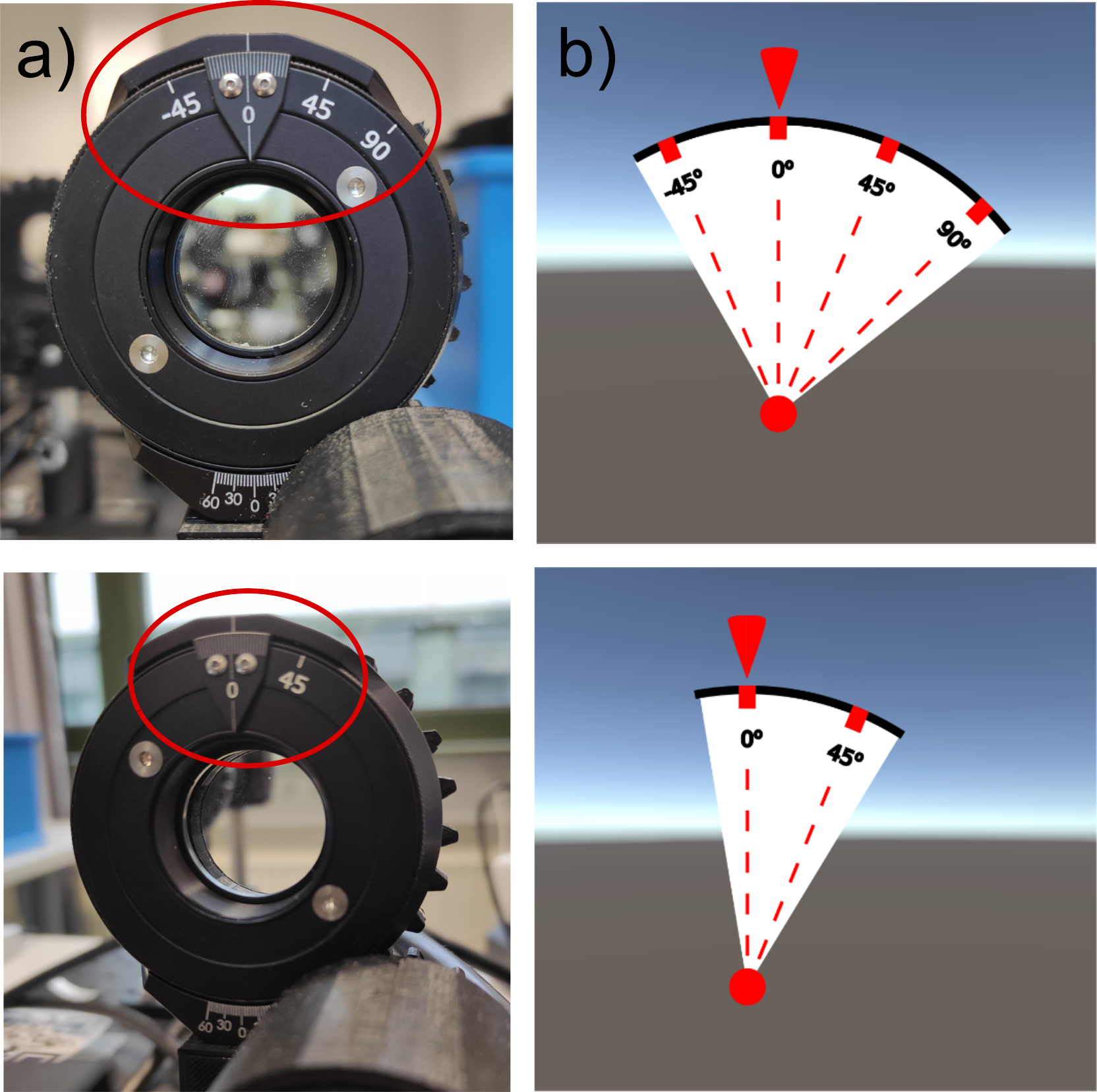}
    \caption{Representations for the $\lambda/2$ plate states in the AR-enhanced quantum cryptography experiment. Column a) shows the phyical markings on the optical instrument itself while column b) depicts the virtual visualizations for the same. The top row contains the representations for the $\lambda/2$ for qubit generation, thus, with the 4 angle positions the respective qubit states can be created. In the bottom row we see the representations for the qubit measurement unit with which one can switch between measuring in basis + or X.}
    \label{fig:lambda_2_repr}
\end{figure}

In the experiment we have two representations for the state of the $\lambda/2$ plates, as shown in Fig. ~\ref{fig:lambda_2_repr}: one is physically engraved notches and values on the respective optical elements and the other is an AR-visualization of a dial showing the current angle with which the plate rotates the polarization of an incoming horizontally polarized qubit. In case of the latter visualization, the sector-shaped dial turns so that the current angle of polarization rotation is shown under the stationary red indicator unit.

Among the key MER functions specified in the DeFT framework, the most applicable one here is a \emph{constraining function} where the interpretation of one representation is constrained by another. While in the virtual visualization the learner sees the values change as they rotate the birefringent crystal, looking at the physical markings on the optical element itself they may notice that the actual angle of rotation for the crystal is half the value for the rotation of the polarization plane. Thus, the physical representation constrains the interpretation of the angle values, dispelling any confusion about what they refer to. In addition, this also prompts the learner to think about the physical process of birefringence and how it relates crystal rotation and polarization plane rotation.

\subsubsection{Qubit states}

In the basic experimental setup we get no direct information about how the state of the qubit evolves from generation up to detection. Being able to follow the evolution of the quantum state of the qubits throughout the experiment is a source of valuable information which also has the potential to encourage the understanding of the underlying physical processes through inquiry-based learning.

\begin{figure}[h!]
    \centering
    \includegraphics[width=0.5\textwidth]{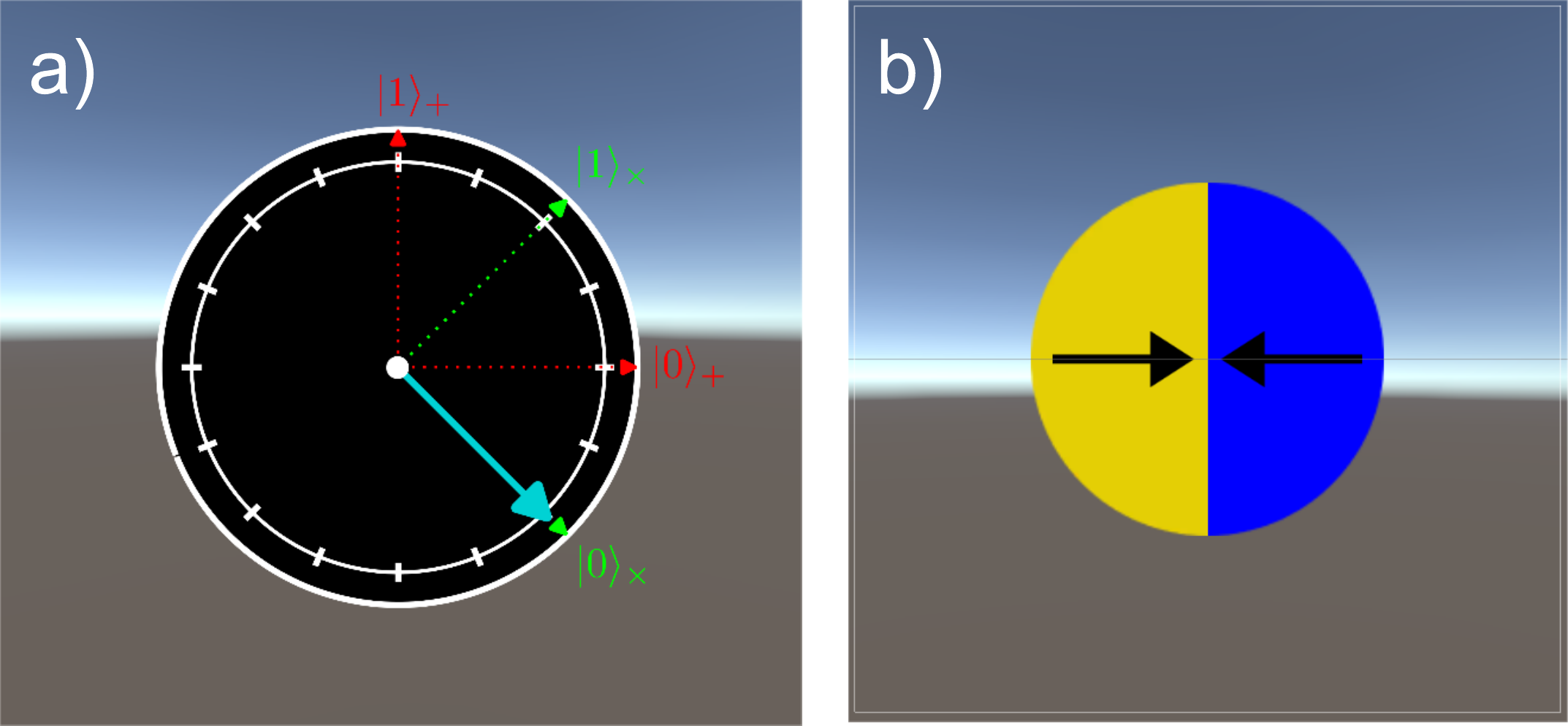}
    \caption{Qubit state representations in the AR-enhanced quantum cryptography experiment. a) The vectorial representation depicts the actual quantum state with a teal vector, the red and green dotted arrows represent the states of the + and X bases, respectively. b) The qake ("Quantum cake") representation shows the quantum state in the + basis, the yellow and blue areas representing the measurement probabilities of the basis states, and the angle of the respective arrows compared to the horizontal represents the phase associated with the coefficients in the coherent superposition of the basis states. Both representations depict the same state, namely, $|\psi\rangle = \frac{1}{\sqrt{2}}(|0\rangle_+ - |1\rangle_+)$}.
    \label{fig:qubit_repr}
\end{figure}

Fig. ~\ref{fig:qubit_repr} shows the two different quantum state representations available in the AR-enhanced experiment: the vectorial representation and what we call the qake ("quantum cake") representation.

The vectorial representation shows the current quantum state as a vector, in relation to the states of the + and X bases. The thick teal vector represents the qubit state and the dotted red and green vectors with the respective labeling show the respective basis states.

The qake ("Quantum cake") model expresses the current state of the qubit in a chosen basis - in our case in the +, i.e., the horizontal/vertical polarization basis. The colors of the two circle segments represent the two states of this particular basis, respectively. The magnitude of the colored areas indicates the probability of measuring the corresponding basis state, i.e., the absolute square value of the associated probability amplitude. The little arrow in each colored area shows the phase of the respective probability amplitude of the quantum superposition state. Thus, a general qubit state can be written as
\begin{equation}
    |\psi\rangle = e^{i\phi_0}\sqrt{A_0}|0\rangle + e^{i\phi_1} \sqrt{A_1}|1\rangle,
\end{equation}
where $A_0$ and $A_1$ are the areas associated with the respective colors, and $\phi_0$ and $\phi_1$ are the counter-clockwise angles of the respective arrows compared to the horizontal direction.

As for the function of these representations within the DeFT framework, it is twofold. They have a \emph{complementary function} because, according to the information they provide. More directly related to the physical realization of the qubit states, the vectorial representation is more suited to be used when carrying out the steps of the QKD protocol than the qake model. On the other hand, these MER also have a \emph{constructing function} by \emph{abstraction}: proceeding from the more particular vectorial to the more abstract qake representation the learner can get a deeper, more general understanding of quantum state superposition that can be transferred to new, not polarization-related quantum systems.

\subsubsection{Research question}

Our main focus is to quantify and compare the visual attention of users on virtual and non-virtual elements of the learning environment during our pilot study, while they are answering questions about Quantum Key Distribution (QKD). To this end, using collected eye gaze data, we calculate the respective total fixation durations and conduct a statistical analysis (a dependent t-test) to determine whether there is a significant difference between the average total fixation durations on virtual and non-virtual elements. We also endeavour to relate the content of the questions to the observed eye gaze behavior to find out what type of content directs the attention to the virtual representations and to the non-virtual elements, respectively.

\section{Methods}\label{sec:eyegaze_analysis}

\subsection{Experimental setup}

\begin{figure}[hbt!]
    \centering
    \includegraphics[width=0.5\textwidth]{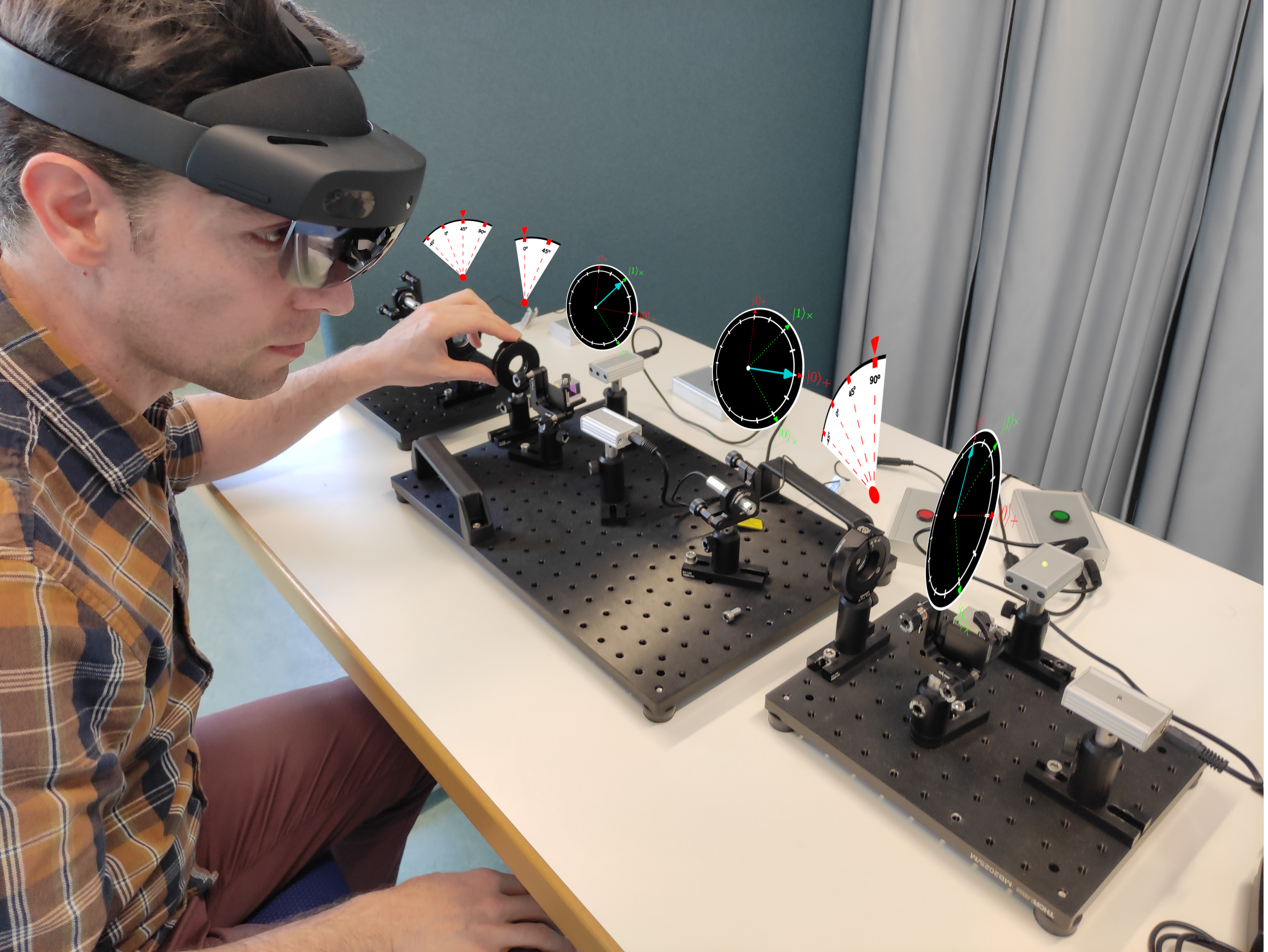}
    \caption{Illustration of the learning environment in use. The virtual visualizations are salient, always turning towards the learner to provide a better user experience.}
    \label{fig:setup_in_use_illust}
\end{figure}

\begin{figure}[hbt!]
    \centering
    \includegraphics[width=\textwidth]{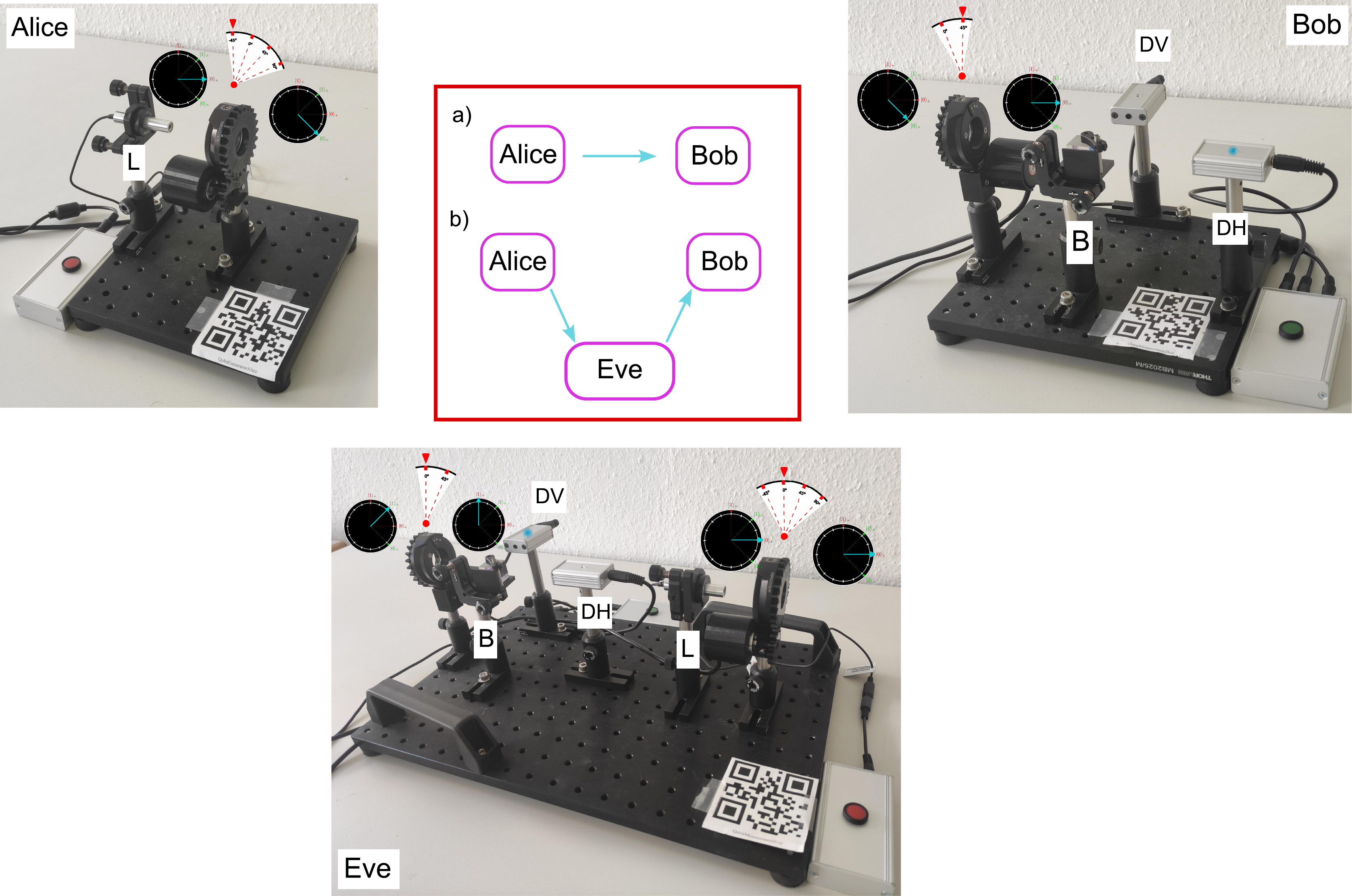}
    \caption{The three units of the experimental setup for generating and transmitting the qubit (Alice), for receiving it (Bob) and for trying to intercept the transition (Eve). They can be combined as shown in a) and in b), respectively. One can see the virtual visualizations for $\lambda/2$ plate settings and qubit states. For the physical components, the notation is as follows. L: laser, B: beamsplitter, DH: detector for horizontally polarized light, DV: detector for vertically polarized light.}
    \label{fig:experimental_setup}
\end{figure}

For our AR-enhanced quantum cryptography student experiment we use the EDU-QCRY1 Quantum Cryptography Demonstration Kit by ThorLabs as hardware. The AR-enhancement is built on top of this physical setup, the corresponding application running on a Microsoft HoloLens2 wearable AR device with eye tracking capability. Fig.~\ref{fig:setup_in_use_illust} shows the learning environment in use as the optical elements are being calibrated and Fig.~\ref{fig:experimental_setup} depicts the full system with detailed explanation of the respective elements of the setup.

An important feature of the enhancement of the basic experiment is that each $\lambda/2$ plate has a rotary encoder attached that is connected to an Arduino device, communicating with the AR app via Bluetooth. This makes it possible for the virtual representation to display the current angle in real time and provide an immersive user experience.

The components of the experiment can be grouped into two groups: transmitter and receiver. 

    A transmitter has a horizontally polarized light source (chosen as the reference angle of $0^o$) realized by a diode laser passing through an initial polarizer and a $\lambda/2$-plate which is used to rotate the light polarization into any of the 4 required basis states.

    A receiver's function is to measure the incoming photons in the + or the X basis, as the user desires. It contains a polarization beam splitter that discriminates between horizontal ($0^o$) and vertical ($45^o$) polarizations. At each output of the beam splitter we have a detector - this allows for measuring the photonic basis states in the + basis. To make it possible to measure the states of the X also, the incoming light passes through a $\lambda/2$ plate prior to hitting the beam splitter. The $\lambda/2$ plate has two pre-defined settings: $0^o$ and $45^o$, allowing the receiver unit to measure the incoming light polarization in the + and X basis, respectively.

    The experimental setup is divided into three units: Alice, Bob and Eve. Alice's setup is a transmitter, Bob's is a receiver while Eve (the eavesdropper) has both: Eve's receiver measures the incoming light from Alice and her transmitter generates a light pulse for Bob to receive. Since the breadboards are separated, by rearranging them it is easy to create a scenario with or without an eavesdropper.

It is important to mention that this is an analogy experiment which means that the single-photon sources are substituted by diode lasers producing short light pulses and the quantum measurement effects are simulated by an electronic module inside the detection apparatus. For instance, when measuring a horizontal polarization in the X basis, both detectors will detect light, but the built-in electronics will randomly assign one of the detectors to click, thus simulating a quantum measurement of state $|0\rangle_+$ in the X basis.

Such an analogy setup has several advantages in a student laboratory course. The absence of single-photon sources makes the setup considerably more affordable, the assembly and alignment of the setup much quicker and easier, furthermore, the hardware is much more robust and portable than a setup with real single-photon generation. In addition, the simpler setup lets students concentrate on the actual objectives of the experiment, without lengthy adjustments of the light sources.

\subsection{Participants and procedure}

We have conducted a pilot study with N=38 students who had to use our learning environment. Each group consisted of 1-3 people, thus, when performing the quantum key distribution, they had to flexibly distribute the roles of Alice, Bob and Eve among themselves.

During the study, the only available visualization for the qubit state was the vectorial representation, since this visualization aligned better with the curriculum of the students. 

\begin{figure}[hbt!]
    \centering
    \includegraphics[width=\textwidth]{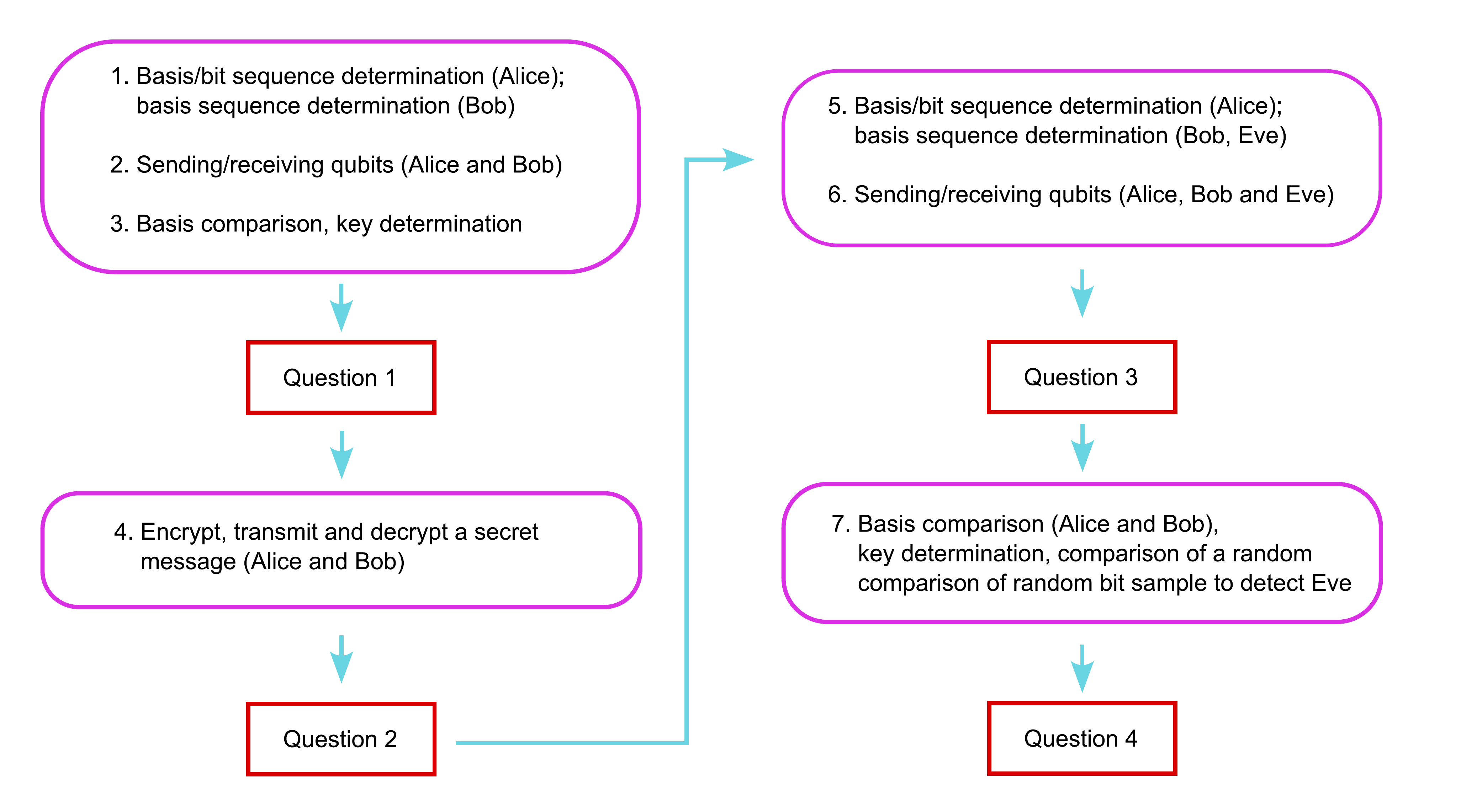}
    \caption{Task sequence in the pilot study. Questions 1, 2, 3 and 4 are included to check the understanding of the participants about quantum cryptography in general and the experiment in particular.}
    \label{fig:study_design}
\end{figure}

The experimental design is shown in Fig. ~\ref{fig:study_design}. Between tasks 3 and 4, 4 and 5, 6 and 7 and after 7, respectively, the learners had to answer a question that tested their knowledge related to QKD in general and the experiment in particular. The first question referred to why in quantum cryptography two bases are used. The second question was about criteria for generating a key. The third question was about the procedure Eve uses to try to intercept the transmission, and the fourth question was about how to detect the presence of an eavesdropper by comparing the bits and bases. 
While pondering the answer, the students were allowed to look at the AR-enhanced experimental apparatus at their leisure and to discuss with each other before giving an answer. 

\subsection{Analysis}
The eye movements of the participants were tracked during the experiment, providing valuable data about cognitive processes during the lab work. This is a novel and so far rarely seen approach to evaluating the attention of students using a AR-enhanced student experiment. The gaze data is provided by the built-in eye tracker of the Microsoft HoloLens2 and gathered by the AR app using the ARETT software package \cite{Kapp2021}. According to the measurement results provided in the paper, the mean sampling time for the eye tracker is $33.33\mathrm{ms}$ (SD $2.5\times 10^{-4}$ms). Table ~\ref{tab:ET_measures} shows the accuracy and precision of the eye gaze data collected using ARETT on a Microsoft HoloLens2.

\begin{table}[h!]
\centering
\begin{tabular}{|l|l|l|}
\hline
Distance to target & Accuracy(SD) in cm/degrees  & Precision(SD) in cm/degrees \\ \hline
0.5 m              & 0.91 (0.41) / 1.00 (0.44)     & 0.40 (0.16) / 0.29 (0.13)                             \\ \hline
1.0 m              & 1.56 (0.83) / 0.85 (0.46)     & 0.67 (0.24) / 0.25 (0.11)                        \\ \hline
2.0 m              & 2.85 (1.31) / 0.77 (0.35)     & 1.35 (0.49) / 0.24 (0.10)                   \\ \hline
4.0 m              & 5.03 (2.27) / 0.68 (0.31)       & 3.12 (1.26) / 0.28 (0.12)         \\ \hline
\end{tabular}
\caption{Accuracy and precision values of mobile eye tracking with ARETT when using the Microsoft HoloLens2 \cite{Kapp2021}, for resting calibration targets at different distances. Our measurements lie in the range of 0.5 m - 1.0 m.}
\label{tab:ET_measures}
\end{table}

For comparing visual attention on virtual and non-virtual elements, we chose to analyze the gaze data from the phases of the experiment where learners are answering questions 1, 2, 3 and 4. We chose these periods because during all the other phases the participants assume diverse roles and, since their respective environments differ, the collected gaze data would be harder to compare. 

While answering the respective questions, the visual attention of each participant is distributed between virtual and non-virtual elements of the learning environment. When comparing them, we are examining a dataset of matched pairs,  thus, the dependent samples t-test is an ideal method to check whether the sample averages of total fixation durations differ significantly.


\section{Results}

\subsection{Visual attention: Descriptive data}
\begin{figure}[h!]
    \centering
    \includegraphics[width=0.9\textwidth]{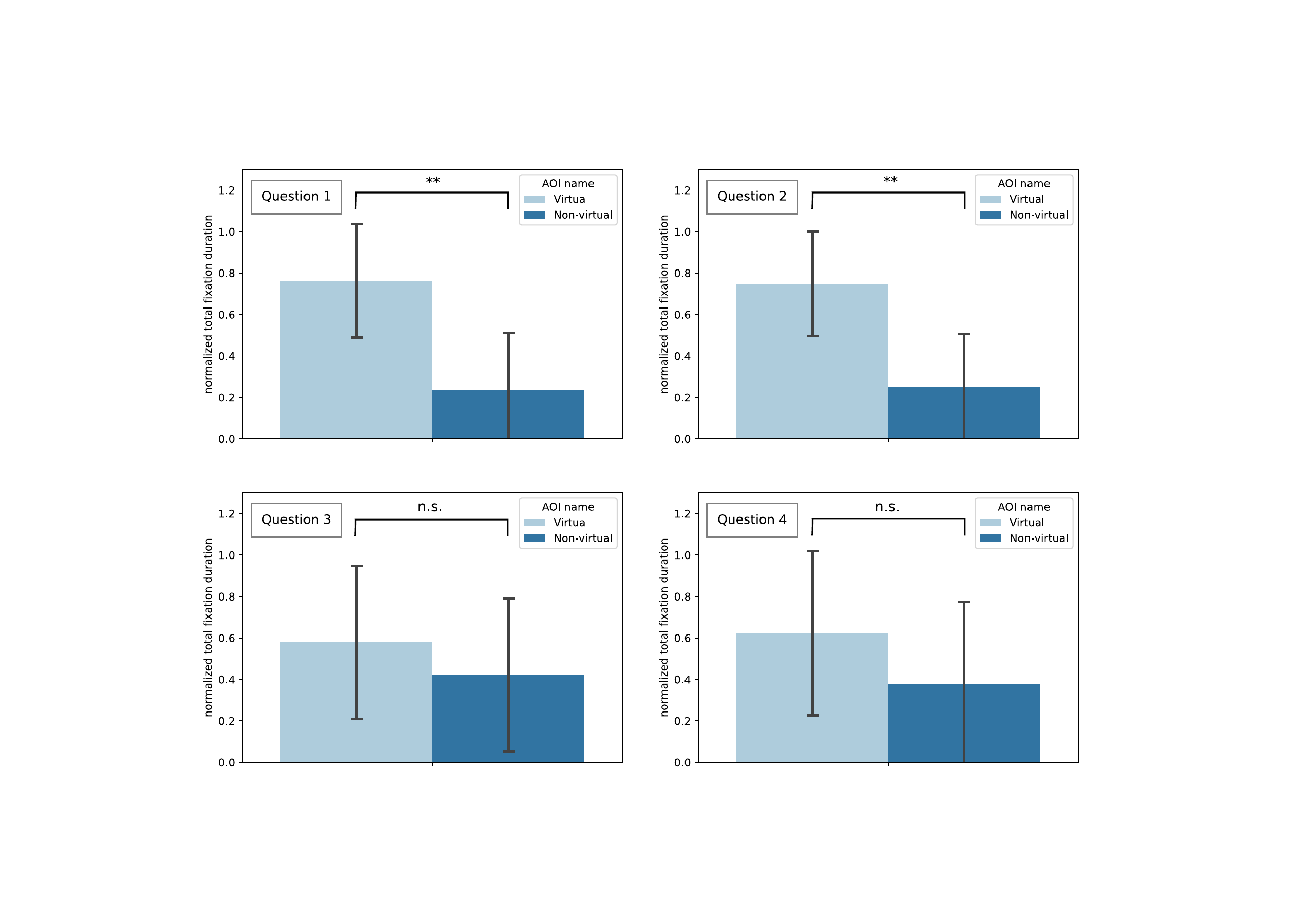}
    \caption{Normalized total fixation durations for virtual and non-virtual (real) elements in the AR-enhanced student experiment for the phase where students answer the 4 questions about the quantum key distribution protocol. The normalization is done by dividing the total fixation duration in the respective AOI (Area Of Interest) by the sum of the total fixation durations over all AOIs. The error bars represent the standard deviation. The horizontal brackets show the results of respective dependent t-tests: the averages for virtual and non-virtual data differ significantly in case of questions 1  and 2  ("**" means $p\leq\alpha=0.01$), while for 3 and 4 there is no significant difference ("n.s."), i.e, $p>\alpha=0.05$.}
    \label{fig:fixations}
\end{figure}

From Fig. ~\ref{fig:fixations}, one can infer how students' visual attention is distributed between the virtual and non-virtual elements of the experiment while answering the questions. The Figure shows the virtual/non-virtual total fixation durations normalized with the sum of all fixation durations during the respective questions.
%
%

The normalized total fixation durations for the virtual and non-virtual elements of the experiment show that students on average look longer at the virtual visualizations. For questions 1 and 2, this difference is more pronounced than for questions 3 and 4.



\subsection{Visual attention: inferential analysis}



To get a quantitative comparison between the respective averages, we conducted a statistical test. We chose the dependent t-test as method for the comparison. 

Table \ref{tab:dep_ttest_results} shows the results of the analysis. Taking $\alpha=0.05$, we can see that indeed, for Questions 1 and 2 the learners focus significantly longer on virtual than on non-virtual elements ($p<\alpha$) while for questions 3 and 4 this difference is not significant ($p>\alpha$). The potential reason behind this is that the first two questions are more abstract, more conceptual, and one can answer them simply by looking at the virtual representations. For questions 3 and 4, the average total fixation duration values are still higher for the virtual objects but the difference between virtual and non-virtual averages is no longer significant. Since questions 3 and 4 refer directly to the experimental steps in the QKD protocol, it stands to reason that the learners look at both the virtual and the physical components. 

It is important to point out that taking into account that we perform 4 measurements (one for each question) so that for a more rigorous consideration, $\alpha$ needs to be divided by 4 (Bonferroni correction): $\alpha_B=0.0125$. However, the result for significant difference holds up for this case as well, being $p<0.0125$ for questions 1 and 2.

\begin{table}[]
\begin{tabular}{|l|ll|ll|l|l|l|}
\hline
\multicolumn{1}{|c|}{Question} & \multicolumn{2}{c|}{Virtual}                          & \multicolumn{2}{c|}{Non-virtual}                      & \multicolumn{1}{c|}{t-statistic} & \multicolumn{1}{c|}{p} & \multicolumn{1}{c|}{Cohen's d} \\
\multicolumn{1}{|c|}{}         & \multicolumn{1}{c|}{Mean}   & \multicolumn{1}{c|}{SD} & \multicolumn{1}{c|}{Mean}   & \multicolumn{1}{c|}{SD} & \multicolumn{1}{c|}{}            & \multicolumn{1}{c|}{}  & \multicolumn{1}{c|}{}          \\ \hline
1                              & \multicolumn{1}{l|}{0.7631} & 0.2638                  & \multicolumn{1}{l|}{0.2369} & 0.2638                  & 3.4547                           & 0.0048                 & 0.9973                         \\ \hline
2                              & \multicolumn{1}{l|}{0.7478} & 0.2440                  & \multicolumn{1}{l|}{0.2522} & 0.2440                  & 3.7988                           & 0.0019                 & 1.0153                         \\ \hline
3                              & \multicolumn{1}{l|}{0.5800} & 0.3550                  & \multicolumn{1}{l|}{0.4210} & 0.3550                  & 0.7715                           & 0.4553                & 0.2227                         \\ \hline
4                              & \multicolumn{1}{l|}{0.6231} & 0.3812                  & \multicolumn{1}{l|}{0.3768} & 0.3812                  & 1.1195                           & 0.2848                 & 0.3232                         \\ \hline
\end{tabular}
\caption{Results of dependent t-test performed on the normalized total fixation durations on virtual and non-virtual elements of the learning environment while learners were solving the respective questions. Both for $\alpha=0.05$ and for the Bonferroni-corrected $\alpha_B=\alpha/4=0.0125$, the p-value for questions 1 and 2 is smaller, so the difference between the averages is significant.}
\label{tab:dep_ttest_results}
\end{table}

Overall, from the collected gaze data, we can see that the virtual visualizations indeed have a strong salience and are extensively being used by learners when engaging with the learning environment. 

\section{Conclusion} \label{sec:conclusion}


This study explores the integration of multiple external representations (MER) in a augmented reality (AR)-enhanced quantum cryptography student experiment using the DeFT (Design, Functions, Tasks) framework. The theoretical framework emphasized how visualizations can promote conceptual learning in quantum physics through three key functions: complementary, constraining, and constructing. Quantum mechanics concepts, often viewed as abstract and mathematical, demand sophisticated representations to make the underlying physics more accessible and comprehensible to learners.

The AR-enhanced environment, featuring both virtual and physical representations of quantum phenomena such as qubit states and measurement processes, was designed to encourage students to explore and reflect on these abstract concepts. The DeFT framework provided a structure to analyze how these representations aid understanding. The virtual visualizations complemented the physical setup by offering immediate feedback and interactive elements, while the constraining function contributed to clarifying ambiguities that might arise from physical representations alone. Finally, the constructing function supported students to build deeper, transferable knowledge of quantum states.

The results of the pilot study, which involved 38 students, provided valuable insights into learner attention, measured through eye-tracking data. Statistical analysis revealed that students generally focused more on virtual representations, especially when answering conceptual questions. Questions requiring a deeper understanding of abstract quantum concepts led to significantly longer fixation durations on virtual elements, indicating their salience in supporting conceptual learning. In contrast, tasks involving more procedural steps in the quantum key distribution (QKD) process showed a more balanced focus between virtual and non-virtual elements.

In conclusion, the use of AR-enhanced experiments combined with a well-structured MER framework such as DeFT offers a promising approach to improving conceptual understanding in quantum physics. The ability to visualize complex quantum phenomena in real time and interact with them in an immersive environment proved beneficial in guiding learner attention and fostering inquiry-based learning. Further research and larger-scale studies could help refine these techniques and expand their application in physics education.
\newpage

\section*{References}

\bibliographystyle{unsrt}
\bibliography{bibliography}

\end{document}